\documentclass[aps,prl,reprint,floatfix,amsmath,amssymb,
superscriptaddress, groupedaddress]{revtex4-1}

\usepackage{graphicx}
\usepackage{epstopdf}
\usepackage{color}
\usepackage{hyperref}
\usepackage{bm}
\usepackage{printlen}
\usepackage{units}
\usepackage{booktabs}
\usepackage{float}
\floatstyle{plaintop}
\restylefloat{table}

\hypersetup{hidelinks}
\renewcommand{\vec}[1]{\mathbf{#1}}

\begin{document}

\title{Origin of  band flatness and  constraints of higher Chern numbers}

\author{Alexander  Kruchkov}   

\affiliation{Department of Physics, Harvard University, Cambridge, MA 02138, USA}    
\affiliation{Department of Theoretical Physics, University of Geneva, Geneva CH 1211, Switzerland}   
\affiliation{Branco Weiss Society in Science, ETH Zurich, Zurich, CH 8092, Switzerland}

\begin{abstract}
Flat bands provide a natural platform for emergent electronic states beyond Landau paradigm. Among those of particular importance are flat Chern bands, including bands of higher Chern numbers ($C$$>$$1$). We introduce a new framework for \textit{band flatness through wave functions}, and classify the existing isolated flat bands in a "periodic table" according to tight binding features and wave function properties. 
 Our flat band categorization encompasses seemingly different classes of flat bands ranging from atomic insulators to perfectly flat Chern bands and   Landau Levels.  The perfectly flat Chern bands satisfy Berry curvature condition $ F_{xy} = \text{Tr} \, \mathcal G_{ij}$ which on the tight-binding level is fulfilled only for infinite-range models. Most of the natural Chern bands fall into category of $C=1$; the complexity of creating higher-$C$ flat bands is beyond the current  technology.  This is due to the breakdown of the microscopic stability for higher-$C$ flatness, seen atomistically e.g. in the increase of the hopping range bound as $\propto$$\sqrt{C} a$. Within our new formalism, we indicate strategies for bypassing higher-$C$ constraints and thus dramatically decreasing the implementation complexity.
\end{abstract}

\maketitle

Electronic band flatness is a new condensed matter paradigm.   The flat bands had been studied in the context of Landau levels \cite{Landau1930, *Landau1977, Haldane2018,Kapit2010}; 
  however a conceptually new approach in material science is engineering flat bands from the first principles \cite{Miyahara2005, Maimaiti2017, Rhim2019}. To date, we have accumulated sufficient numerical evidence of flat bands formation ranging from artificial crystallographic lattices (Kagome, Lieb) to controllable flat band engineering in twisted van der Waals heterostructures \cite{Trambly2010,Naik2018,Carr2020}. The controllable and predictable engineering of band flatness is of strategical importance, especially for strongly correlated phases such as unconventional superconductivity \cite{Cao2018,Yankowitz2019,Park2021,Hao2021}, Fractional Chern Insulators \cite{Neupert2011, Tang2011, Sun2011, Haldane2011}, and the SYK phase \cite{Kitaev2015,Sachdev1993,Chen2018}, to name a few. Of particular importance are flat Chern bands, however the higher-order Chern numbers are a  rare case in reality, with the majority of natural flat bands  restricted to  $|C|=1$.

\renewcommand*{\arraystretch}{1.2}
\begin{table*}
\begin{ruledtabular}
\caption{"Periodic table" for perfectly flat gapped bands categorized through criteria \eqref{flatness}, \eqref{flat1}, \eqref{flat2}. \vspace{2 mm}}
\begin{tabular}{p{0.5 in} p{0.6 in} p{0.6 in} p{0.7 in} p{0.7  in} p{0.7  in} p{0.7  in} p{0.75 in} p{0.7 in}  p{0.7 in}}
atomic insulator &  fine-tuned flat & generic trivial   nonlocal & \multicolumn{3}{c}{generic topological  nonlocal}   & Landau Level & TBG chiral  & \\
\hline
0  &  $\mathcal O (1)$ &  $\infty$ & $\infty$ & $\infty$ & $\infty$ & $\infty$  & ($\infty$); $\mathcal O (1)$  &  \textbf{Hopping range} $\Lambda$\\
-  &   none  &  any  &  $\sim \pi/a$  & $\sim \pi/a$  & $\sim \pi/a$  & -  & - (cancelled; $\pi/\lambda_M$) &  \textbf{Singularity position}  $h$\\
0  &  0 &  0 & $|C| =\frac{1}{2} + \frac{1}{2}$ & $|C| = m$ & $|C|>1$  & $C= \pm 1$  & $C$=$\pm 1$  &  \textbf{Chern number}  $C$\\
{\footnotesize{not defined}}  &  {\footnotesize{double-periodic, nonholomorphic}} &  {\footnotesize{double-periodic, nonholomorphic}}  & {\footnotesize{double-periodic nonholomorphic}} & {\footnotesize{meromorphic non-double-periodic}} & {\footnotesize{double-periodic meromorphic}} &  {\footnotesize{holomorphic quasiperiodic}}  & {\footnotesize{holomorphic quasiperiodic}}&  \textbf{Periodicity in BZ and analyticity} \\
\end{tabular}
\end{ruledtabular}
\end{table*}

It is certainly surprising that in many cases we can point out on the common origin for perfectly flat bands, which is (self)-trapping in a limited coordinate region. 
In case of Landau levels with $\psi_{\text{LL}}$$\sim$$e^{-r^2/2 l_B^2}$, the self-trapping occurs in the area $\sim$$l_B^2$; for atomic insulator the self-trapping occurs on the lattice sites; for fine-tuned flat-bands (such as in Kagome, Liebe lattices) the self-trapping happens within the plaquette $\sim$$a^2$; in twisted van der Waals heterostructures the self-trapping happens at small part of the moir{\'e} cell around high-symmetry points (AA stacking in twisted bilayer graphene). This motivates us to define the \textit{band flatness parameter} through wave functions
\begin{align}
\vspace{- 1 mm}
f  =  \frac{ \sum_{R > \Lambda} |\Psi (R) |^2  }{ \sum_{R > 0} |\Psi (R) |^2 }.   
\label{flatness} 
\end{align}
\noindent
Here $\Lambda$ is a system-dependent real space hopping range (dimensionless). The parameter $f$$\ll$$1$  is  identifying  band flatness  through wave function localization.  Clearly $\delta(R)$ gives a perfectly flat band with $f$$=$$0$, and the flatness construction at small $\Lambda$$\sim$$1$ is intuitively clear.  
 Let's illustrate that definition \eqref{flatness} works also for $\Lambda \to \infty$, both for the gapped trivial and topological bands.

\textit{Generic construction of flat bands.}---Consider a generic tight binding Hamiltonian $\mathcal H = \sum_{ij}^{\Lambda} t_{ij}^{\alpha \beta} c^{\dag}_{i \, \alpha} c^{\dag}_{j \, \beta}$. Here $\Lambda$ is the hopping range. Setting here lattice constant $a$$\to$$\infty$ (or $t_{ij}$$=$$0$) immediately recovers the "atomic insulator", a system with a perfectly flat band at $E_0$$=$$0$.  The atomic insulator is the simplest perfectly flat band opening the "periodic table" of perfectly flat bands (Table I). We can systematically build other classes of flat bands from the tight binding.  In the momentum representation the generic tight-binding Hamiltonian reads $\mathcal H  = \sum_{\vec k} c^{\dag}_{\vec k \alpha} \mathcal  H_{\alpha \beta} (\vec k) c^{}_{\vec k \beta} $.  The matrix Hamiltonian $\mathcal  H_{\alpha \beta} (\vec k)$ has electronic bands  $\varepsilon_1(k), \varepsilon_2(k) ... \varepsilon_N(k)$. Among them, we pick up  the $n$$^\text{th}$ band which is dispersive, gapped out from the others, and  not crossing zero. We further introduce a new Hamiltonian hosting the \textit{perfectly flat band} through procedure

\vspace{-7 mm}
 \begin{align}
t_{ij} \  \ \to \ \ 
 T^{\textbf{flat}}_{ij} = \frac{E_0}{N}  \sum_{\vec k } 
 \frac{ \mathcal  H (\vec k) }{\varepsilon_n(k)} e^{i \vec k (\vec R_i - \vec R_j) } . 
 \label{flat}
 \end{align}
   \vspace{-3 mm}
 
\noindent
The new tight binding Hamiltonian with hopping \eqref{flat} $\mathcal H^{\textbf{flat}} = \sum_{ij}^{\infty}  T^{\textbf{flat}}_{ij} \, c^{\dag}_{i \, \alpha} c^{\dag}_{j \, \beta}$ contains at least one perfectly flat band positioned at $E= E_0$.  However, this unique flat band comes at the expense that the tight-binding model becomes nonlocal ($\Lambda'$$\to$$\infty$).  Of course, one can now take only the first few terms of $T^{\textbf{flat}}_{ij}$, since they are decaying exponentially fast with distance. However any real space truncation of $\Lambda$ inevitably introduces a "parasitic dispersion", so the band is no longer perfectly flat. The analysis above is useful for understanding the "fine tuning" flat bands in Kagome, Lieb and other lattices. In the rare cases, one can find a lucky choice of fine-tuning parameter set upon truncation to nearest neighbors NN and NNN ($\Lambda$$\approx$$2,3$), which gives flat or perfectly flat band through destructive interference. However, this construction is  a fine tuning and inclusion of further-order hopping generically breaks flatness.

Clearly, the same band flattening procedure \eqref{flat} can be applied to the Chern bands, as far as they are  gapped. In this case the hopping parameters are generically complex-valued. For storytelling, consider the Haldane model \cite{Haldane1988} on NN and NNN hoppings ($\Lambda =2$), 
\begin{align}
\mathcal H_{\text{FDMH}}  = \sum_{i} t_0 c^{\dag}_i c^{}_i + \sum_{\langle ij \rangle} t^{\text{NN}}_{ij} c^{\dag}_{i } c^{\dag}_{j } +  \sum_{\langle\langle ij \rangle \rangle} t^{\text{NNN}}_{ij} c^{\dag}_{i } c^{\dag}_{j },
\end{align}
\noindent 
where we fix $t^{\text{NN}}_{ij}$ as real and $t^{\text{NNN}}_{ij}$$=$$t'e^{i \Phi_{ij}}$. Clearly, for $\Phi_{ij}$$=$$\pm \pi/2$, the spectrum becomes particle-hole symmetric \cite{Haldane1988}. This yields upon transformation (1)  two pefectly flat bands positioned at $E$$=$$\pm E_0$. The resulting $\mathcal H_{\text{FDMH}}^{\textbf{flat}} $ becomes nonlocal ($\Lambda'$$=$$\infty$), though it is possible to truncate it and make the band arbitrary flat by choosing a corresponding truncation $\Lambda'$ and minimizing bandwidth (see e.g. Ref.\cite{Neupert2011}). 
Thus it is always possible to construct a perfectly flat  band for $\Lambda$$=$$\infty$, be it topological or not. 
Since the range $\Lambda$ of effective tight-binding is related to the wave function's spatial tails, the above statement is consistent with the flatness parameter \eqref{flatness}.

\textit{Band flatness for topologically-trivial bands}. Addressing localization properties is more natural in terms of Wannier functions.
 Topologically-trivial flat bands are maximally Wannierizable, thus 
without loss of generality we consider localization along one of the two spatial dimensions and investigate its asymptotics \cite{Kohn1959,He2001}

\vspace{-4 mm}
\begin{align}
\mathcal{W}(x - R) \propto  \int d k_x \, e^{i k_x (x-R)} \, u_{k}  . 
\label{Wannier}
\end{align}
\vspace{-3 mm}

\noindent
A topologically trivial band may have a singularity in the form of a branch vertex in complex momentum $k = k_x + i h_x$, of a generic form
$u (k \approx k_*) \simeq u_0 [i (k-k_*)]^{\alpha} $, with $\alpha$$>$$-1$. Denote $k_*= k_0+ ih$ as the position of the singularity in the complex plain. Hence  the asymptotics of \eqref{Wannier} reduces to integral representation of the Gamma function along a Hankel contour encompassing $k_0+i h$. The result for large $x$ is 
\begin{align}
\mathcal{W}(x) \simeq 2 u_0 \sin (\pi \alpha) \Gamma(1+ \alpha) \frac{\exp(-h x)}{x^{1+\alpha}}, 
\end{align}
\noindent
In case of several singularities we take $h = \text{min}[ \, \text {Im} \, k_*]$, corresponding to the one closest to the real axis.  We further use this asymptotic to derive the behavior of the flatness parameter 
\eqref{flatness} as

\vspace{-6 mm}
\begin{align}
f  \sim \frac{1}{\Lambda^{2 (\alpha+1)} } e^{-2 h \Lambda a} , 
\label{flat-triv}
\end{align}
\noindent
For this, we need to use analytical estimates of the sums of form $\Sigma_n (\Lambda)$$=$$\sum_{x=\Lambda}^{\infty}  x^{-n} e^{-x}$. We rewrite this sum as $ \Sigma_n (\Lambda)$$=$$e^{- \Lambda} \, \zeta (\frac{i}{2 \pi}, n , \Lambda)$ where $\zeta (\phi, n , \Lambda) = \sum_{x=0}^{\infty} (x+\Lambda)^{-n}e^{2 \pi i \phi x}$ is Lerch zeta function (see e.g. \cite{Apostol1951,*Johnson1974}). Up to $\mathcal O (1)$ prefactor Lerch zeta function $\zeta (\frac{i}{2 \pi}, n , \Lambda)$ behaves as $1/\Lambda^n$ for $\Lambda$$\gg$$1$, thus we obtain analytical estimate  $\Sigma_n (\Lambda)$$\sim$$\Lambda^{-n} e^{-\Lambda}$ and  $\Sigma_n (1)$$\sim$$\mathcal O (1/e)$. The flatness parameter involves summation  of form $\Sigma_n  (\Lambda +1)/\Sigma_n (1)$$\sim$$(\Lambda+1)^{-n} e^{- \Lambda} $. We confirm numerically that this estimate works good even for $\Lambda$$\sim$$2$. Taking now $n = 2 (\alpha +1)$ and restoring dimensional parameters, we obtain \eqref{flat-triv}. 
For our purposes, we are not interested here in the power-law prefactor, and factor of 2 in the exponent. It is safe to rewrite the flatness criterion as 
\begin{align}
 f_0  = e^{- h \Lambda a }, \ \ \ \text{for trivial bands. }
\label{flat1}
\end{align}
\noindent
The flatness parameter $f_0$ of \eqref{flat1} sets a fundamental scale for achievable band flatness, and covers three distinguished classes of \textit{perfectly flat} nontopological bands with $f_0 = 0$: 
\vspace{-1 mm}
\begin{itemize}
\item $a$$\to$$\infty$, atomic insulator. 
\vspace{-1 mm}
\item $\Lambda$$\to$$\infty$, generic flat band construction, Eq. \eqref{flat}. 
\vspace{-1 mm}
\item $h$$\to$$\infty$, singularity removed to infinity (nonsingular perfectly flat band). 
\vspace{-1 mm}
\end{itemize}
\noindent 
The first two cases are discussed on page 1; the examples for nonsingular perfectly flat bands  are listed in Ref. \cite{Rhim2019}. Cases of topologically trivial, gapped perfectly flat bands  are covered by the three classes above, and constitute the topologically-trivial sector of the flat band classification (Table I). We do not have evidence of  perfectly flat, gapped topologially-trivial bands which do not fit into this classification. We now proceed to the Chern bands.

\textit{Band flatness for Chern bands}. We start from the  construction of higher-$C$ Chern bands on the basis of double periodic meromorphic functions. The essential toolbox is built upon implementation of theta functions, Weierstrass and Jacobi functions, and their combinations \cite{Akhiezer1990}. Independently of the choice of the elliptic function, we can use connection between the wave function singularities (poles) and the band Chern number $C$$=$$\int_{\text{BZ}} \frac{d^2 \vec k}{2 \pi} F_{xy}$, with $F_{xy}$$=$$\partial_x A_y$$-$$\partial_y A_x$,  $\vec A_{\vec k}$$=$$- i \langle u_{\vec k} | \partial_{\vec k} u_{\vec k} \rangle$), in the complex plane $z$$=$($k_x$,$k_y$) \cite{Jian2013,Baum1970}

\vspace{-5 mm}
\begin{align}
C= \left[ \oint_{\gamma_{\text{BZ}}} +  \sum_{z^*_i} \oint_{\gamma_{z_i^*}}  \right] \frac{ A_{\bar z} dz + A_{z} d \bar z }{4 \pi}
= \sum_{z^*_i} p_i (z_i^*). 
\label{Chern}
\end{align}
The Chern number is expressed through the sum of all poles $z^*_i$ in Brilloin zone (BZ), counting their multiplicity $p_i (z^*_i)$ \cite{Baum1970}. 
The definition \eqref{Chern} is important since it is not connected to the band dispersion as such.

A theorem, tracing back to Thouless \cite{Thouless1984}, prevents Wannierizing Chern bands in 2D; see recent discussion in \cite{Po2018}. However, as was pointed out by Qi, it does not prevent Wannierizing a Chern band along one of the 1D directions of a 2D Chern insulator \cite{Qi2011}.   
(Moreover, in this way a duality between a $C=1$ flat Chern band and the Lowest Landau level is established \cite{Qi2011}). 

For our purposes, localization along 1D is a good indicator of band flatness through \eqref{flatness}, \eqref{Wannier}. We thus proceed with Wannierizing a Chern band along 1D, and finding its asymptotic behavior. 
For this, it is sufficient to replace the elliptic functions with their principal behavior around poles
\begin{align}
 u(k) \simeq  \sum_n \frac{u_0}{[i (k - k^*_n)]^{p_n}}  + \text{Regular part}.
\label{meromorphic}
 \end{align}
 (here $k = k_x+i k_y$ is in the first BZ). The associated Chern number is $C = \sum_n p_n$. The main contribution to the integral \eqref{Wannier} is given by the pole \eqref{meromorphic} of multiplicity $p_n \leq C/2$  closest to the the real axis.  The residue at the pole is $\text{Res} \, u (k) = -i u_0 x^{p_n-1} e^{i x k_*}/(p_n-1)!$, with $k_* = k_0 + i h$. Using the residue theorem, and an appropriate contour, one obtains asymptote   \begin{align}
\mathcal{W}(x) \simeq \frac{2 \pi u_0}{(p_n-1)!} x^{p_n-1} e^{- h x}.
\label{Wannier-top}
\end{align}
To derive the flatness parameter, we need to evaluate sums of form $\Sigma_m (\Lambda) = \sum_{x > \Lambda}^{\infty} x^m e^{-x}$ with $m$$=$$2 (p_n-1)$. We can rewrite this sum through Lerch zeta function as $\Sigma_m (\Lambda) = e^{-\Lambda} \zeta(\frac{i}{2 \pi}, -m, \Lambda)$.  Up to $\mathcal O (1)$ prefactor Lerch zeta function $\zeta (\frac{i}{2 \pi}, -m , \Lambda)$ behaves as $\Lambda^m$ for $\Lambda$$\gg$$1$, thus  $\Sigma_m (\Lambda)$$\sim$$\Lambda^{m} e^{-\Lambda}$ and  $\Sigma_m (1)$$\sim$$\mathcal O (1/e)$. Restoring dimensional units, we obtain the flatness criterion as  
$
f \sim \Lambda^{m} e^{- 2ha \Lambda}
$, 
where $m = 2(p_n-1)$, i.e. depends on the nature of the wave function singularities.  We confirm numerically that this is a good fit \footnote{Numerical analysis gives $[(\Lambda+ \Lambda_1)/\Lambda_2]^{m} e^{- 2 ha\Lambda}$ with $\Lambda_1 \approx \Lambda_2$.}.

\begin{figure}[t]
\includegraphics[ width = 1.0 \columnwidth]{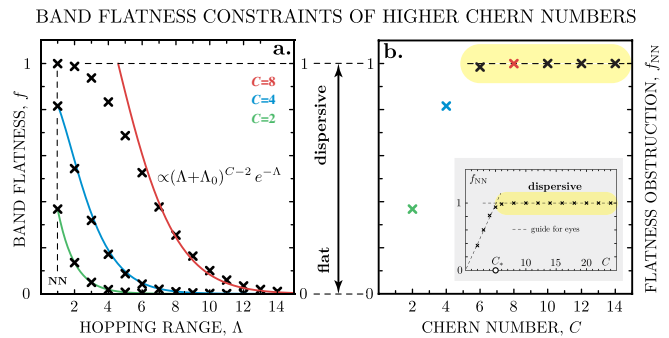}
\caption{\textbf{Flatness of  higher-Chern bands and topological obstructions}. \textbf{(a)} Band flatness bound of \eqref{flatness} vs. electronic hopping range $\Lambda$ for Chern bands with $C=2,4,8$. Solid line represents asymptote \eqref{flat-top1} [31]; $2ha$$=$$1$ is set for all data points.  Chern bands can be perfectly flat only for non-local hopping ($\Lambda$$\to$$\infty$). \textbf{(b)} Flatness bound at NN hoping represents obstruction to band flatness such that inclusion of further-order hoppings is required (Fig.2). In case of Chern bands, $f_{\text{NN}}(C)$ quickly saturates to the maximally-obstructed at $C_*(h)$ (Fig.1b); in case of trivial bands $f_{\text{NN}}$ lower bound can be set to zero \eqref{flat1}.  This explains why most of natural flat Chern bands are limited to unit $|C|$=$1$. }
\end{figure}

We show that the finite Chern number inevitably leads to constraints on the band flatness (Fig.1).  We note that high $C$$=$$N$ in \eqref{Chern} can be attained in multiple ways. The simplest way is by having two poles of multiplicity $N$ each \footnote{Double periodic meromorphic function cannot have less than two zeros,  two poles counting multiplicity}. This results into a higher Chern number  $C_{2N}$$=$$2 N$$\gg$$1$ restraining band flatness as $
f$$\sim$$\Lambda^{C_{2N}-2} e^{- 2ha \Lambda}$. 
To have a topological band, the wave function singularity must reside inside the BZ. This leads to the limitation $h$$\leq$$\pi/a$ (square lattice), or $h a$$\sim$$1$ \textit{independently} of  $a$ and lattice symmetries. The flatness parameter is 

\vspace{-5 mm}
\begin{align}
f \sim \Lambda^{C_{2N}-2} e^{- \Lambda}.  
\label{flat-top1}
\end{align}
\noindent
The only way to have a perfectly flat Chern band ($f$$=$$0$) is for $\Lambda$$\to$$\infty$. We come to the same conclusion if we consider a higher $C$$=$$N$ Chern band constructed through adding $N$ poles of multiplicity $1$ (or any other combination). In this case we need to replace $C_{2N}$ with $2$$\times$$C_1$,  $C_1$$=$$1$ in \eqref{flat-top1}, however the same argument holds: the Chern band is perfectly flat only for $\Lambda$$\to$$\infty$. We cannot fine-tune local tight-binding with NN and NNN hoppings in order to have perfectly flat Chern band on the lattice, as we did in the topologically-trivial case. The band flatness constraints in this case results from the inevitable singularity of the wave functions of form $1/(k-k_*)^\alpha$, which is irremovable by definition, and is the source for the Berry curvature. 

Thus we come to the conclusion that the Chern bands cannot be perfectly flat unless in the case of infinite hopping range $\Lambda$. 
This observation is consistent with Chen theorem (2014), stating that in the double-periodic system the perfectly flat Chern bands are accessible only for $\Lambda = \infty$ \cite{Chen2014}.

\textit{Flat Chern bands with C=1}.---Now we can relax some of the conditions imposed in Eqs. \eqref{meromorphic}, \eqref{Chern} on the flat band wave function. This can be done in two ways: by relaxing double periodicity or by relaxing holomorphicity conditions. 
 First, we can relax condition of double-periodicity, but still require the flat band state to be a function of $z$$=$$k_x$$+$$i k_y$.
 In this way we unlock all the odd Chern numbers in \eqref{Chern}, including unitary $|C|=1$. In this case the contribution along the BZ boundary, which vanishes due to double-periodicity in \eqref{Chern}, may itself contribute to the Chern number. Interestingly, we can go even further by canceling the remaining singularity(-ies) inside the BZ and producing an effective magnetic flux; in this case the Chern number is given purely by the circulation along the BZ boundary $\oint_{\gamma} \vec A d \vec k = 2 \pi C$ \cite{Zak1989}. This case corresponds to the continuum model of twisted bilayer graphene (TBG), which hosts perfectly flat Chern bands at the magic angle \cite{SanJose2012,TKV}. Several authors has pointed out on the duality between the perfectly flat Chern bands in TBG and the lowest Landau level; we refer to Refs.\cite{SanJose2012,TKV,Popov2021} for further details. In both cases we are dealing with effective magnetic fields which produce Berry curvature $F_{xy} \propto l_B^2$, flux $2 \pi$ through effective Brillouin zone (MBZ) and a perfectly flat band in a certain limit. Following \cite{Qi2011}, without loss of generality we can consider asymptote $\mathcal W (x) \propto x^n e^{-x^2/2 l_B^2}$. Thus for Landau Levels the flatness criterion \eqref{flatness} asymptotically reads 
 
 \vspace{-5 mm}
 \begin{align}
 f_{\text{LL}} \sim \Lambda^{2n -1} e^{-\Lambda^2 a^2/l_B^2}, 
 \end{align}
where we assume $a$$\ll$$l_B$.  We see that the Landau levels are perfectly flat only in the nonlocal limit 
$\Lambda$$\gg$$l_B/a $$\to$$\infty$. 
Bringing this system on the tight-binding lattice (finite $a$, finite $\Lambda$) inevitably broadens the Landau levels for any finite $\Lambda$, see Refs.\cite{Hofstadter1976,Kapit2010,Dong2020}.

We can further relax the condition of holomorphicity, but keep the condition of double periodicity. This correspond to the Chern bands constructed from the pairs of Dirac points. A simplest construction is $C=1/2+1/2$ Chern band constructed from two Dirac cones carrying Chern number $|C|=1/2$ each. The total Chern number is integer since the Dirac nodes always come in pairs on the lattice. 
  Since the normalized wave function associated with a (gapped) Dirac cone contains singularity of form $u (k)$$\sim$$(k-k_*)^{-1/4}$, the Wannier asymptotics is of form $\mathcal W(x)$$\sim$$x^{-3/4} e^{-h x}$. Since we again have restriction $h a$$\sim$$1$, we arrive to the flatness criterion of  form $f \sim \Lambda^{-3/2} e^{- \Lambda}$. Once again we cannot have a perfectly flat band at local tight binding. Note that we can also construct higher Chern number $C=N$ by taking at least $2 N$ Dirac cones and melting them into a flattened band (a strategy proved to be fruitful in graphene multilayers). This case completes our discussion on flatness of Chern bands, since further dropping both the condition of periodicity and holomorphicity breaks down the concept of Chern number as an integer topological invariant on the tight-binding level.

Independently of the method the flat Chern band has been composed, it is safe to rewrite the flatness criterion as 

\vspace{-7 mm}
\begin{align}
f_0 = \Lambda^{S[C]} e^{- F(\Lambda)}, \ \ \ \text{for topological bands},
\label{flat2}
\end{align}
\noindent
where $S[C]\propto|C|$, $F(\Lambda)\propto \Lambda, \Lambda^2$ with system-dependent coefficients which reflect the way the Chern number $C$ is constructed.  The higher Chern number is, the harder is to make a good flat band.   
The topological bands can be perfectly flat only at infinite hopping range.

Comparing criterion \eqref{flat2} with the $f_0$ criterion for trivial bands \eqref{flat1}, we notice that criterion \eqref{flat2} is not reducible to the atomic insulator. This is because the Chern insulator and atomic insulator belong to different topological classes, and cannot be adiabatically connected \cite{Chiu2016}.

\textit{Microscopic analysis and hopping range bounds.}---We now look deeper into the microscopic details of topological constraints (Fig. 2).  
A representative parameter is the hopping range scale, defined for an isolated band as

\vspace{-6 mm}
\begin{align}
r_{\text{hop} } = \left( 
\frac{ \sum \limits_{i,j=0}^{\infty}  (x_i - x_j)^2 \mathcal{W} (x_i) \mathcal{W} (x_j)     } {  \sum \limits\limits_{i,j=0}^{\infty}  \mathcal{W} (x_i) \mathcal{W}(x_j)   }   \right)^{1/2} .
\nonumber
\end{align}
For definiteness, we use the construction of higher-Chern flat band as in Eqs.(\ref{Wannier-top}-\ref{flat-top1}). We observe numerically that $r_{\text{hop}}^2$ behaves monotonically with $\tilde h$$=$$h a$; for $h a$$\lesssim$$1$ we have $r_{\text{hop}}^2$$\sim$$1/\tilde h^2$; $r_{\text{hop}}^2$$\sim$$\,C$.   
To understand this asymptotics analytically, we replace sum with integrals in definition above, $r_{\text{hop}}^2$$\simeq$$h^{-2} I[p_n$$-$$1,2]/I[p_n$$-$$1,0]$, with $I[q,s]$$=$$\iint \limits_{0}^{\infty} dx_1 dx_2   (x_1 x_2)^q (x_1$$-$$x_2)^s e^{- (x_1+x_2)} $. We arrive to the analytical expression $r_{\text{hop}}^2$$=$$2 \Gamma(p_n+1)/h^2 \Gamma(p_n)$. Since in this case $p_n$ are integers, we have $r_{\text{hop}}^2$$=$$2 p_n/h^2$. Restoring connection between the poles and the Chern number as in Eq.\eqref{flat-top1}, we  obtain

\vspace{-5 mm}
\begin{align}
r_{\text{hop} } \simeq \frac{1}{h} \sqrt {C} .  
\label{rhop1}
\end{align}
\noindent 
Hence the higher-$C$ bands require longer-range hopping.

The wave function singularity position  $h$=$h_{\text{max}}$  results into  the \textit{lower bound} for hopping range \eqref{rhop1}. For the Chern bands, it is impossible to remove this singularity to infinity, thus $h_{\text{max}}$ is bounded from above. For a square lattice, the first estimate on $h_{\text{max}}$ gives as $\pi/a$. Independently of lattice symmetries, we can use $h_{\text{max}} a$$\sim$$1$, hence the lower bound for hopping range in case of topological bands scales as (Fig.2a)

\vspace{-5 mm}
 \begin{align} 
r_{\text{hop} } \simeq \sqrt {C} a.  
\label{rhop2}
\end{align}
This agrees with Ref.\cite{Jian2013}. Note that the actual hopping range cutoff $\Lambda_*$ required to stabilize the band of desirable flatness $f_*$ with respect to next-order hoppings can be  much higher than the lower bound \eqref{rhop2} scaling as $\propto$$Ca$  (Fig.2b).

\begin{figure}[t]
\includegraphics[ width = 1.0 \columnwidth]{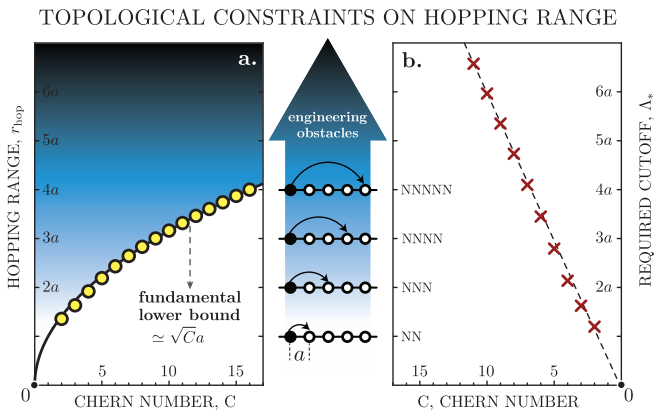}
\caption{\textbf{Microscopical analysis of topological constraints on the band flatness.} \textbf{(a)} Characteristic hopping range scale $r_\text{hop}$ demonstrates the lower bound $\simeq$$\sqrt C a$, Eqs.(\ref{rhop1}-\ref{rhop2}). This lower bound is fundamental \cite{Jian2013}. \textbf{(b)}  Required tight-binding cut-off $\Lambda_*$ to stabilize band flatness $f_*$$\ll$$1$ is higher than the hopping range lower bound $r_0$. For all data points, we take $ha$$=$$1$ and $f_*(\Lambda_*)$$=$$0.1$. For higher-Chern flat bands, both  \textbf{(a,b)} require precise experimental control of multiple hopping parameters \textit{beyond} NN and NNN, thus restricting realizations of $|C|$$\gg$$1$ flat bands. The strategy to bypass this limitation is proposed in the main text. }
\end{figure} 

\textit{Bypassing higher Chern constraints.}---We notice that the microscopic constraints of form \eqref{rhop1},\eqref{rhop2}, which were derived assuming higher-order poles of multiplicity $N$$=$$C/2$, could be bypassed in the experiment. Instead, we can take (a physically reasonable number) $N$ of simple poles to construct $C$$=$$N$$\times$$C_1$ flat band using $C_1$$=$$1$. In this case $\mathcal W (x)$$\sim$$x^{C_1 -1} e^{-h a x}$, hence we need to replace $C$ with $C_1$ in \eqref{rhop2}, and the lower bound $\sqrt{C_1}$$a \sim a$ remains fixed on the lattice size for all the Chern numbers.    Consider a thin film material (e.g. a monolayer or bilayer) with its band structure hosting a flat Chern band of unit $|C_1|$$=$$1$. We can  stack $N$ such layers into a multilayer heterostructure, represented by matrix Hamiltonian $\mathcal H$ 
with monolayer Hamiltonians on the main diagonal 
and interlayer coupling terms 
placed on the diagonals just below and above. 
The topological invariant of this composition can be found through computing 
$
C[ \mathbb G ]$$=$$\frac{1}{3!} \sum_{ijk} \epsilon_{ijk} \text{Tr} \, \mathbb G  \partial_i \mathbb G^{-1} \mathbb G \partial_j \mathbb G^{-1}  \mathbb G \partial_k \mathbb G^{-1} ,
\label{multiChern}
$
where $i,j,k$$=$$\omega$,$k_x$,$k_y$ and $\mathbb G = (\omega - \mathcal H)^{-1}$ is  matrix Green's function \cite{Gurarie2011}.  
 Consider a case of vanishingly weak interlayer interaction; in this case the Hamiltonian approximately factorizes into matrix tensor product. If the Chern number, associated with a single layer is $C_1$, the $N$ layers in this case give
 \begin{align}
 C[ \mathbb G_N ] = N \times C_1,  \ \ \ 
 \to 
 \ \ \ 
r_{\text{hop} } \sim a. 
\label{rhop3}  
 \end{align}
 This argument is expected to be valid when the weak interlayer interactions are switched on while preserving the finite gap; a similar algorithm was implemented by  Trescher and Bergholtz  \cite{Trescher2012}. Importantly, the construction \eqref{rhop3} does not obstruct the band flatness, i.e. the flat Chern bands can be constructed for any number of layers $N$. Clearly, this  argument is also valid for constructing higher Chern numbers by bringing together $2 n$ (gapped) Dirac cones in van der Waals multilayers for reaching $C$$=$$1,2,3...$ nearly flat Chern bands \cite{Zhang2019}.

\textit{Quantum geometry and flatness constraints}.---Finally, we make connection between the band geometry and flatness criterion \eqref{flatness}.   The band topology \textit{and} geometry is described by the "quantum geometric" tensor \cite{Roy2014,Jackson2015}:  
\begin{align}
\mathfrak{G}_{ij} = \langle \partial_i u_{\vec k} |  
\left( 
1- {|  u_{\vec k} \rangle   \langle  u_{\vec k} | }
\right)
 |  \partial_j u_{\vec k} \rangle,
\nonumber
\end{align}
\noindent
where $\partial_i = \frac{\partial}{\partial k_i}$. The imaginary part  of $\mathfrak{G}$ is responsible for topology, and gives (off-diagonal) Berry curvature $ F_{ij} = \text{Im} \mathfrak{G}_{ij}$; the real part $ \mathcal G_{ij} = \text{Re} \mathfrak{G}_{ij}$ is Fubini-Study metrics and is responsible for the band geometry and its flatness. 
The ideal flat Chern  bands satisfy the Berry-geometric condition $F_{xy} = \text {Tr} \, \mathcal G_{ij}$ (see Refs.\cite{Haldane2011,Roy2014,Claassen2015}).  The holomorphic (and meromorphic) perfectly flat Chern bands automatically satisfy this criterion %\footnote{Indeed, for $u(z) = (u_1(z), ...u_N(z))$, a direct calculation  gives $F_{xy} = 2 \sum_{n\ne m} |u_m \partial_z u_n - u_n \partial_z u_m|^2/ |u|^4 = \text{Tr} \mathcal \, \mathcal G_{ij}$. } 
\footnote{The lowest Landau level automatically satisfies $F_{xy}$$=$$\text {Tr} \, \mathcal G_{ij}$; a delicate moment here is discussion of the higher Landau levels which at first glance violate  this condition. However, since the wave functions of the higher Landau levels can be chosen holomorphic \cite{Haldane2018,Chen2020}, surprisingly this argument still holds.} \nocite{Haldane2018,Chen2020}.  
 This identity requires $\Lambda = \infty$, thus consistent with our classification of Table I. We can further rewrite 
\begin{align}
 F_{xy} = \text {Tr} \, \mathcal G_{ij} = \langle  u_{\vec k} | \, |\hat{\vec r} |^2  \, |  u_{\vec k} \rangle, 
\label{geometry}
\end{align}
where $\hat r$ is the generalized position operator.  Integrating \eqref{geometry} over Brillouin zone, one obtains $r_0^2$$\propto$$C$, hence the localization length is 
$r_0$$\sim$$\sqrt{C} a$.  
For the Chern bands it is impossible to minimize localization length $r_0$ independently from the hopping range bound \eqref{rhop2}; thus the flatness parameter \eqref{flatness} cannot be made arbitrary small for any finite $\Lambda$, there are always finite tails.    The Chern bands can  be made \textit{perfectly flat} only for $\Lambda = \infty$, as was demonstrated through Eq. \eqref{flat2}. This is an intuitive interpretation of Chen theorem \cite{Chen2014}.  Clearly now, the higher Berry fluxes in \eqref{geometry}, hence the higher Chern numbers $|C|$$>$$1$, present stronger constraints on electronic band flatness.

\

\textit{Conclusions.}---To conclude, the new criterion for band flatness \eqref{flatness} allows us to derive two concise cases $f_0$$=$$e^{-ha \Lambda}$ for trivial bands and $f_0$$=$$\Lambda^{S[C]} e^{- F[\Lambda]}$ for topological bands of Chern number $C$. The criteria allows us to systematize the known classes of perfectly flat bands (Table I) as the fundamental building blocks for realistic nearly flat bands. The large Chern number  presents an obstruction to the band flatness, which can be seen in e.g. in the required range $\propto$ $\sqrt C a$. The most feasible route for overcoming this obstruction is using multilayers for gluing together higher Chern numbers.  
We can hope that the strategy for building higher-$C$ bands with proposed pathways \eqref{rhop3} shall become a route towards realizing elusive Kitaev $E_8$ state, a bosonic Quantum Hall phase which can be build on the $C$$=$$8 n$ flat band \cite{Kitaev2006}.

\begin{center}
$\bigodot$
\end{center}

\newpage

\textit{Acknowledgements}. The author thanks Bertrand Halperin, Subir Sachdev, Philip Kim, and Emil Bergholtz for enlightening discussions. This work was financed by the Branco Weiss Society in Science, ETH Zurich, through the research grant on flat bands, strong interactions and the SYK physics.  

\bibliography{Refs}

\end{document}